\documentclass[twocolumn,aps,prl,showpacs,floatfix,superscriptaddress]{revtex4}
\usepackage[T1]{fontenc}
\usepackage[latin9]{inputenc}
\usepackage{amsmath}
\usepackage{amssymb}
\usepackage{graphicx}
\usepackage{float}
\usepackage{siunitx}
\usepackage[english]{babel}

\makeatletter
\@ifundefined{textcolor}{}
{%
 \definecolor{BLACK}{gray}{0}
 \definecolor{WHITE}{gray}{1}
 \definecolor{RED}{rgb}{1,0,0}
 \definecolor{GREEN}{rgb}{0,1,0}
 \definecolor{BLUE}{rgb}{0,0,1}
 \definecolor{CYAN}{cmyk}{1,0,0,0}
 \definecolor{MAGENTA}{cmyk}{0,1,0,0}
 \definecolor{YELLOW}{cmyk}{0,0,1,0}
}

\@ifundefined{definecolor}{\usepackage{color}}{}

\usepackage{babel}

\makeatother

\usepackage{babel}
\begin{document}
\flushbottom

\title{Self-trapping of exciton-polariton condensates}

\author{Dario~Ballarini}
\affiliation{CNR NANOTEC---Institute of Nanotechnology, Via Monteroni, 73100 Lecce, Italy}

\author{Igor~Chestnov}
\affiliation{Institute of Natural Sciences, Westlake Institute for Advanced Study, Westlake University, Hangzhou, China}
\affiliation{Vladimir State University, 600000 Vladimir, Russia}

\author{Davide~Caputo}
\affiliation{CNR NANOTEC---Institute of Nanotechnology, Via Monteroni, 73100 Lecce, Italy}
\affiliation{University of Salento, Via Arnesano, 73100 Lecce, Italy}

\author{Milena~De~Giorgi}
\affiliation{CNR NANOTEC---Institute of Nanotechnology, Via Monteroni, 73100 Lecce, Italy}

\author{Lorenzo~Dominici}
\affiliation{CNR NANOTEC---Institute of Nanotechnology, Via Monteroni, 73100 Lecce, Italy}

\author{Kenneth~West}
\affiliation{PRISM, Princeton Institute for the Science and Technology of Materials, Princeton Unviversity, Princeton, NJ 08540}

\author{Loren~N.~Pfeiffer}
\affiliation{PRISM, Princeton Institute for the Science and Technology of Materials, Princeton Unviversity, Princeton, NJ 08540}

\author{Giuseppe~Gigli}
\affiliation{CNR NANOTEC---Institute of Nanotechnology, Via Monteroni, 73100 Lecce, Italy}
\affiliation{University of Salento, Via Arnesano, 73100 Lecce, Italy}

\author{Alexey~Kavokin}
\affiliation{Institute of Natural Sciences, Westlake Institute for Advanced Study, Westlake University, Hangzhou, China}

\author{Daniele~Sanvitto}
\affiliation{CNR NANOTEC---Institute of Nanotechnology, Via Monteroni, 73100 Lecce, Italy}
\affiliation{INFN, Sez. Lecce, 73100 Lecce, Italy}

\begin{abstract}
The self-trapping of exciton-polariton condensates is demonstrated and explained by the formation of a new polaron-like state. Above the polariton lasing threshold, local variation of the lattice temperature provides the mechanism for an attractive interaction between polaritons. Due to this attraction, the condensate collapses into a small bright spot. Its position and momentum variances approach the Heisenberg quantum limit. The  self-trapping  does not require  either  a resonant driving force or a presence of defects. The trapped state is stabilized by the phonon-assisted stimulated scattering of excitons into the polariton condensate. While the formation mechanism of the observed self-trapped state is similar to the Landau-Pekar polaron model, this state is populated by several thousands of quasiparticles, in a strike contrast to the conventional single-particle polaron state.
\end{abstract}


\maketitle
\section*{Introduction}
Self-trapping is an intriguing physical effect that is rather rarely encountered in solid state systems. Among its most well-known manifestations are the phonon self-trapping of charge carriers leading to the polaron formation \cite{Landau1948,koschorreck2012} and the magnetic self-trapping of carriers or excitons in diluted magnetic semiconductors leading to the magnetic polaron formation \cite{dietl1982,ryabchenko1983}. Both polaronic self-trapping mechanisms seem incompatible with the Bose-Einstein condensation (BEC) as the latter leads to the formation of spatially extended coherent many-body states: the bosonic condensates. Nevertheless, experimentally, bosonic condensates of half-light-half-matter quasiparticles, exciton-polaritons, formed in semiconductor microcavities \cite{KasprzakBEC} are frequently spatially localized \cite{kasprzak2007}. The localization is due to the stationary disorder potential necessarily present in the microcavity plane \cite{lagoudakis2010}, and/or due to finite size optical spots used to generate the polaritons \cite{ohadi2016}. The interplay and competition between magnetic self-trapping and BEC of exciton polaritons was analyzed theoretically \cite{shelykh2009}. In non-magnetic systems the polariton liquid intrinsically tends to delocalize as the polaritons repel each other and are repelled by the reservoir of non-condensed excitons formed by a non-resonant optical pumping. It comes as a surprise, that in certain experimental conditions, condensates of polaritons collaps in real space leading to formation of dense droplets \cite{dominici2015}. 

In the present work, we report on the experimental evidence for a steady state self-trapped polariton condensate. We observe the unpinned polariton droplet whose area is two orders of magnitude smaller than the area of the excitation light spot. Looking at the blue shift of the condensate energy and the image of the condensate in the reciprocal space, we conclude  on the origin of self-trapping. The shrinkage of the condensate wave-function is induced by the local heating of the crystal lattice in the center of the pump-beam area. The shrinked state of the condensate is stabilized by the non-linearity caused by the condensate-reservoir coupling.

The observed state constitutes a new phase of a bosonic condensate that may be referred to as the bosonic polaron state. In contrast to a conventional (fermionic) polaron, this state is characterized by a local lattice temperature variation rather than by strain. In our system the polaron wave-function represents the localized many-body wave-function of a bosonic condensate. Self-trapping of exciton polaritons is a new phenomenon in the physics of bosonic condensates. It is promising for applications in polariton simulators \cite{lagoudakis2017,berloff2017}.

\section*{Results}

The system under study is a high quality factor semiconducor microcavity (see Methods) characterized by a long polariton lifetime $\tau=\SI{100}{\pico\second}$ and  a strongly reduced density of defects~\cite{Ballarini2017}. Similar samples have been recently made available by the advanced control of the molecular beam epitaxy (MBE) growth process, that allows for the observation of thermalised polariton condensation and for the efficient ballistic propagation of polaritons in the cavity plane~\cite{Sun2017, Caputo2018,Steger2015}. The microcavity is excited by a continuous wave laser, tuned~$\approx\SI{100}{\milli\electronvolt}$ above the exciton resonance, and the photoluminescence (PL) is collected by imaging spectroscopy (see Methods), allowing for two-dimensional (2D) energy-resolved detection of the emission in both the real-space and the reciprocal-space.

\begin{figure}[!h]
\includegraphics[width=1\linewidth]{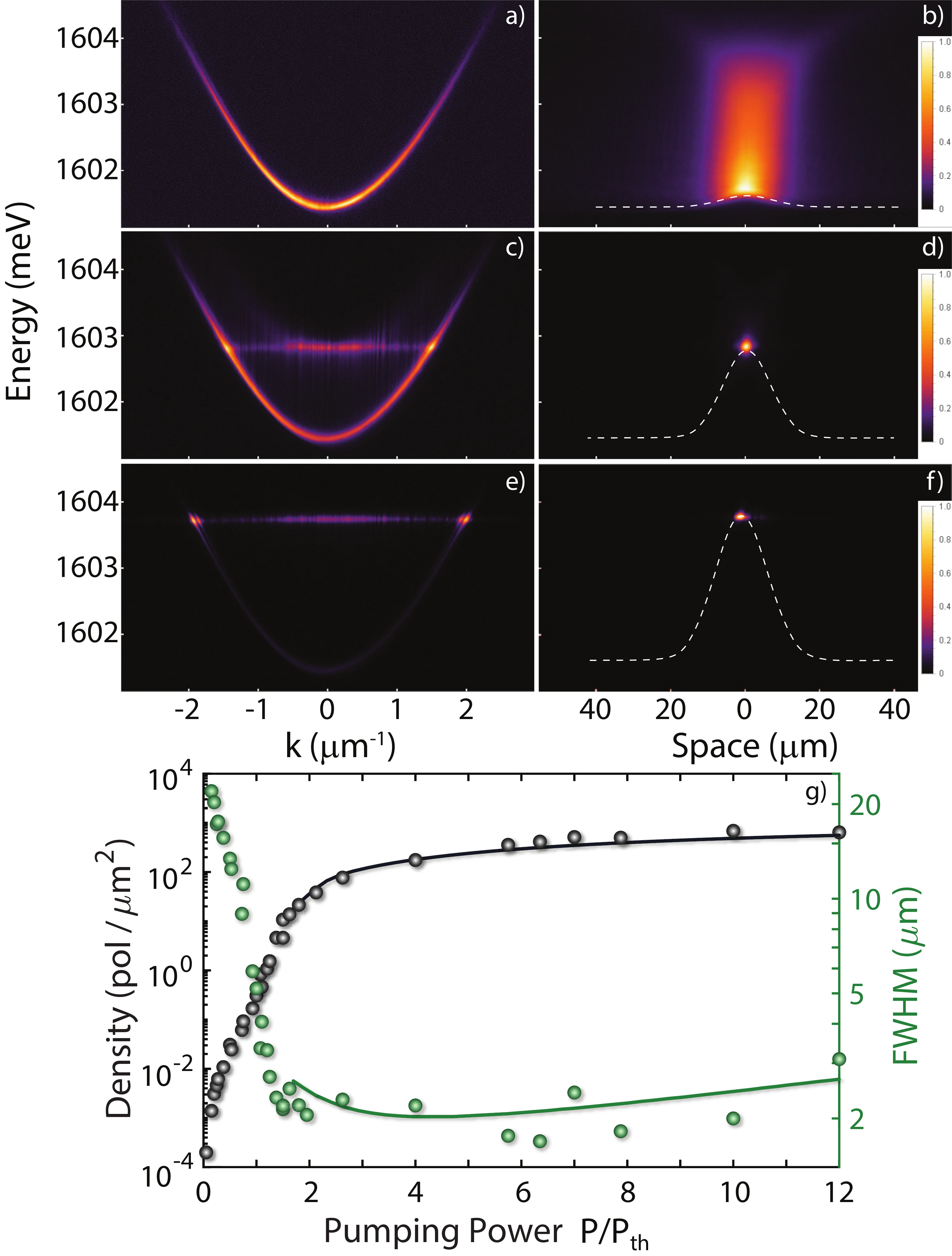} \caption{\textbf{a-f}, Energy resolved (vertical axis) polariton emission in momentum- (left column) and real-space (right column) along the $k_{x}$ and $x$ direction, respectively. The wavevactor $k$ is proportional to the in-plane momentum, $k=p/\hbar=\frac{2 \pi}{\lambda}\sin{\theta}$, with $\lambda$ being the polariton wavelength and $\theta$ being the incidence angle. Pump power increases from $P=0.42P_{th}$, $P=P_{th}$ and $P=3.3P_{th}$ from top to bottom rows. \textbf{g}, Polariton density vs excitation power (black points) normalized at the threshold value (around $P_{th}=\SI{40}{\milli\watt}$). The FWHM (gaussian fit) of the polariton distribution in space is shown (green points) to shrink from values comparable to that of the laser spot (FWHM=$\SI{20}{\micro\meter}$) at low density to less than $\SI{2}{\micro\meter}$ above $P_{th}$. Solid lines are the calculated density (black line) and spatial FWHM (green line) as obtained from numerical simulations (above threshold).}
\label{fig:1}
\end{figure}

In Figs.~\ref{fig:1}a-f, the energy-resolved PL both in momentum (left column) and real space (right column) is shown for increasing excitation powers (from top to bottom). Carrier relaxation leads to the formation of a large population of excitons (exciton reservoir) with in-plane wavevector $k$ beyond the light cone. The exciton relaxation mediated by phonon scattering leads to the occupation of radiative polariton states at the bottom of the lower polariton branch (LPB), close to $k=0$ (see Fig.~\ref{fig:1}a-b). Polariton condensation occurs when the balance between gain (stimulated scattering into the bottom of the LPB) and losses (mainly radiative emission) is achieved, condition that is fulfilled in Fig.~\ref{fig:1}c-d. The emission of the polariton condensate appears blueshifted by $\approx\SI{2}{\milli\electronvolt}$: this is due to the repulsive exciton-exciton interactions in the spatial region where the exciton reservoir is highly populated. The white dashed line in Fig.~\ref{fig:1}b, Fig.~\ref{fig:1}d and Fig.~\ref{fig:1}f denotes the bottom energy of the LPB, that follows the intensity profile of the exciting laser given the short exciton diffusion length ($<\SI{3}{\micro\meter}$). In Fig.~\ref{fig:1}e, the two bright points at $k\approx\pm\SI{2}{\micro\meter^{-1}}$ correspond to polaritons propagating either inwards, bringing the polariton density to the center of the excitation spot, or outwards (in the tail regions) due to the repulsion from the reservoir~\cite{Wertz2010, Kammann2012}. In the same Fig.~\ref{fig:1}e, the central emission with a wide $k$-distribution around $k=0$ corresponds to the spatially localized polariton condensate shown in Fig.~\ref{fig:1}f.

Surprisingly, we found that, even well above the condensation threshold, the emission is localised in a small spatial region of less than $\SI{5}{\micro\meter\squared}$ area. Note that the Gaussian laser spot excites a $10^2$ larger area and the localisation is observed also if changing the position of the excitation spot on the sample, i.e. it is independent on the microscopic details of the sample such as possible defects. The shrinking of the polariton emission above threshold is quantitatively described in Fig.~1g, where the spatial full-width-half-maximum (FWHM) of the polariton emission is shown (green points) for increasing excitation powers and compared to the emission intensity (black points), proportional to the polariton density (see Methods).
\begin{figure}[h]
  \includegraphics[width=1\linewidth]{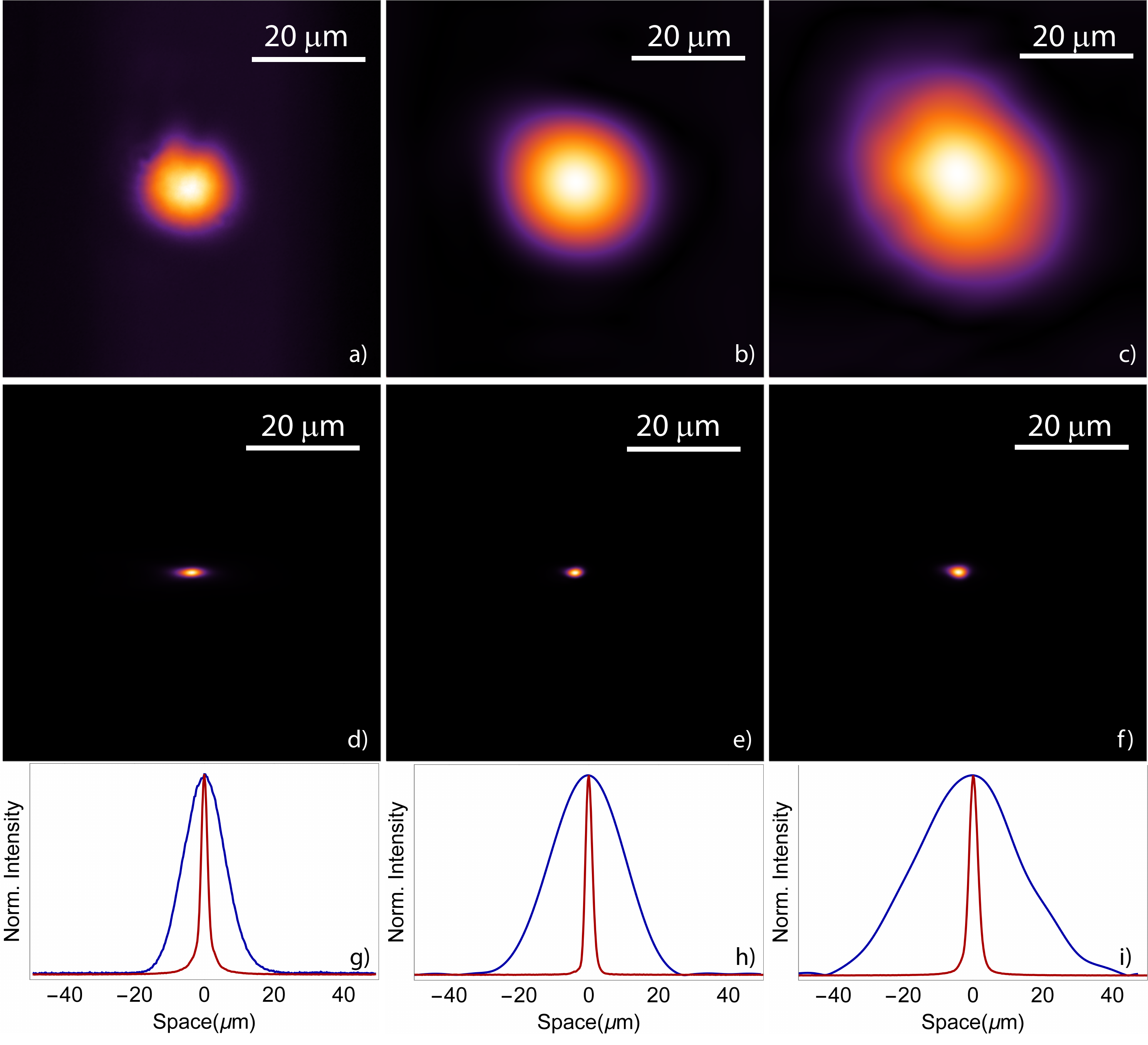}
  \caption{
    \textbf{a,b,c} Three different sizes of the non-resonant pumping laser spot: FWHM is $\SI{20}{\micro\meter}$, $\SI{30}{\micro\meter}$ and $\SI{40}{\micro\meter}$ in \textbf{a}, \textbf{b} and \textbf{c}, respectively. \textbf{d,e,f} Condensate emission in real space corresponding to the excitation spots shown in \textbf{a}, \textbf{b} and \textbf{c}, respectively. The polaron-condensate is always localised in a spot of less than $\SI{5}{\micro\meter}$ diameter. \textbf{g,h,i} The vertical cross-section of the intensity profile of the pumping laser (blue line, corresponding to \textbf{a}, \textbf{b} and \textbf{c}, respectively) and of the condensate (red line, corresponding to \textbf{d}, \textbf{e} and \textbf{f}, respectively).
    }
  \label{fig:2}
\end{figure}

In Fig.~\ref{fig:2}, the emission of the polariton condensate is shown for different sizes of the excitation laser spot (FWHM=20,~30,~40~$\SI{}{\micro\meter}$). The pump power has been adjusted to have the same blueshift of the polariton condensate in all cases despite of the different excitation areas. The full two-dimensional collapse of the polariton condensate is clearly evident by comparing the pump spot intensity (top row) with the PL intensity (second row), showing localised emission from a spatial region two orders of magnitude smaller than the excitation area.
Differently from previous reports, where a confined condensate or the presence of multiple, fragmented condensates have been associated with the presence of the disorder  potential \cite{kasprzak2007,lagoudakis2010}, the high spatial homogeneity of the present sample and the absence of potential defects in the region under study clearly indicate that we observe a purely self-induced localisation.

\begin{figure}[h]
  \includegraphics[width=1\linewidth]{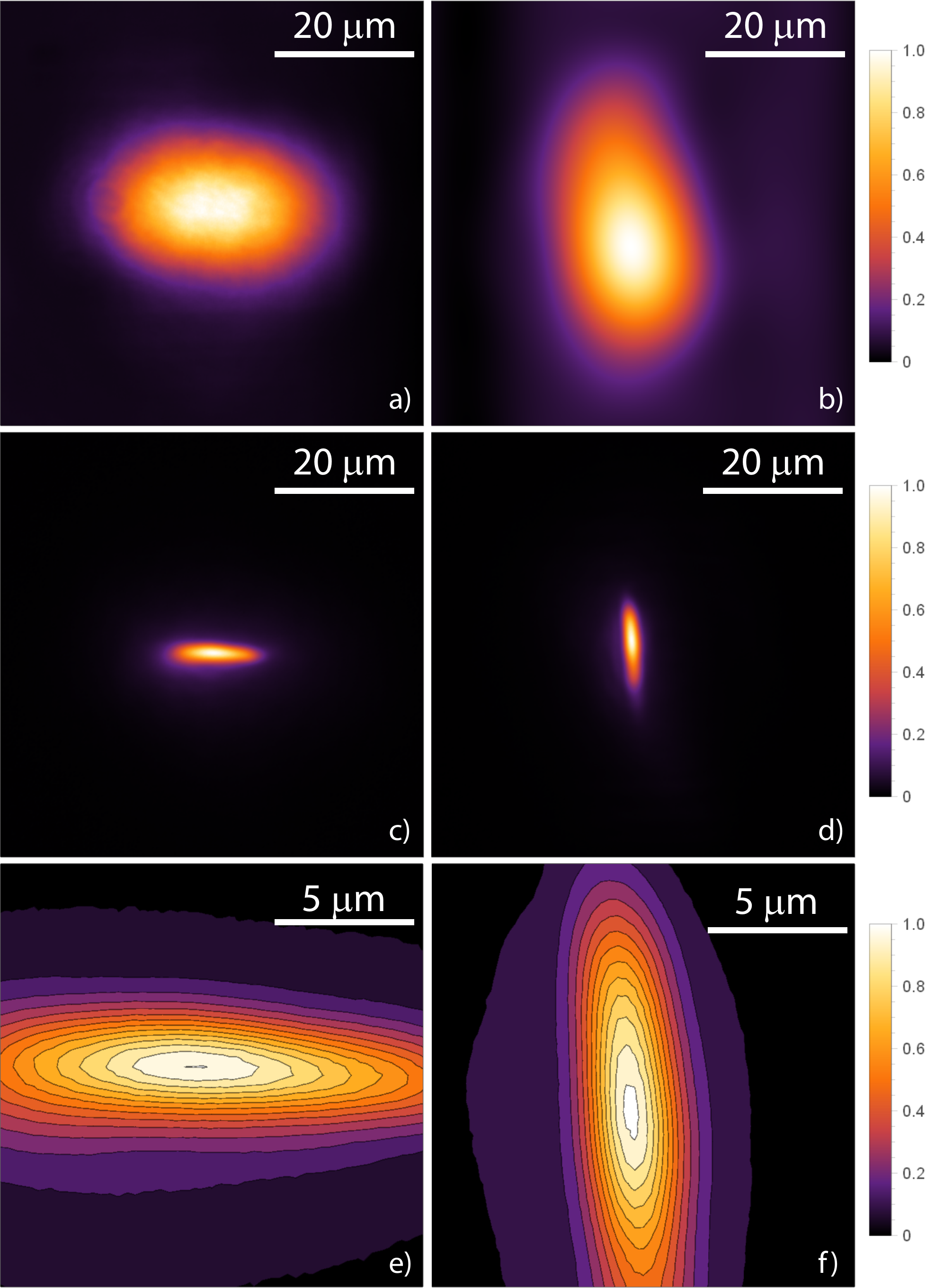}
  \caption{
    \textbf{a,b,} Pumping spots of elliptical shapes with horizontal and vertical elongation, respectively. \textbf{c,d,} Condensate emission in real space corresponding to the excitation shown in \textbf{a} and \textbf{b}, respectively. \textbf{e,f,} Contour plot of the condensate emission intensity (magnification of \textbf{c} and \textbf{d}, respectively).}
\label{fig:3}
\end{figure}

It is interesting to note that, despite of the small size of the localised condensate, its shape can be externally controlled by reshaping the pump beam. To proof this point, the horizontal and vertical elongations of the laser spot have been induced by cylindrical lenses in the optical excitation path (top row in Fig.~\ref{fig:3}). The ellipticity of the shape of the excitation spot is imprinted on the localised polariton condensate, as shown in the central row of Fig.~\ref{fig:3} and in the magnified contour plot of the emission intensity in Fig.~\ref{fig:3}e and Fig.~\ref{fig:3}f. Conversely, the localisation is absent when a top-hat laser spot profile is used (see SI), showing that a gradient of the potential landscape is needed. This suggests that localisation occurs at the position of the highest exciton density, where the condensation threshold is first reached, and it  is feeded by a net flux of polaritons directed towards the localisation point. A similar scenario, with the emergence of a heat-assisted, sink-type steady state, has been recently predicted in one-dimensional polariton wires~\cite{Chestnov2018}.

\section*{Numerical simulations}

The real-space collapse of a polariton condensate indicates the presence  of a self-induced trapping mechanism capable to overcome the intrinsic polariton-polariton repulsion. This requires an effective attractive interaction that cannot be explained in the conventional theoretical framework used for polaritons, that is a generalised Gross-Pitaevskii equation with positive interaction constants. We attribute this attraction to the polariton energy renormalization induced by the heat released during the condensation. Polariton thermalization to the ground state is assisted by the emission of the acoustic phonons which heat the crystal lattice and may localy change the lattice temperature by several degrees \cite{klembt2015}. This leads to a local modification of the polariton dispersion and results in the formation of the trap at the position of the condensate density peak where the temperature is maximal. This trap triggers the real-space collapse of the condensate and its shrinkage into a tight spot. The self-trapped state is further stabilized due to the stimulated scattering of reservoir excitons to the condensate.

To describe our experiments, we use the Ginzburg-Landau equation for the polariton order parameter $\Psi$ coupled to the kinetic equation for the density $n$ of incoherent polaritons (see Methods). The effective attractive interaction is proportional to the rate of polariton relaxation from the reservoir to the ground state which is determined by the product of their densities, $|\Psi|^2 n$. Searching for the stationary solution in the case of a symmetric Gaussian pump, we reproduce the experimentally observed real-space condensate distribution, see Fig.~\ref{fig:2} and Fig.~S1.

\begin{figure}[h]
  \includegraphics[width=0.75\linewidth]{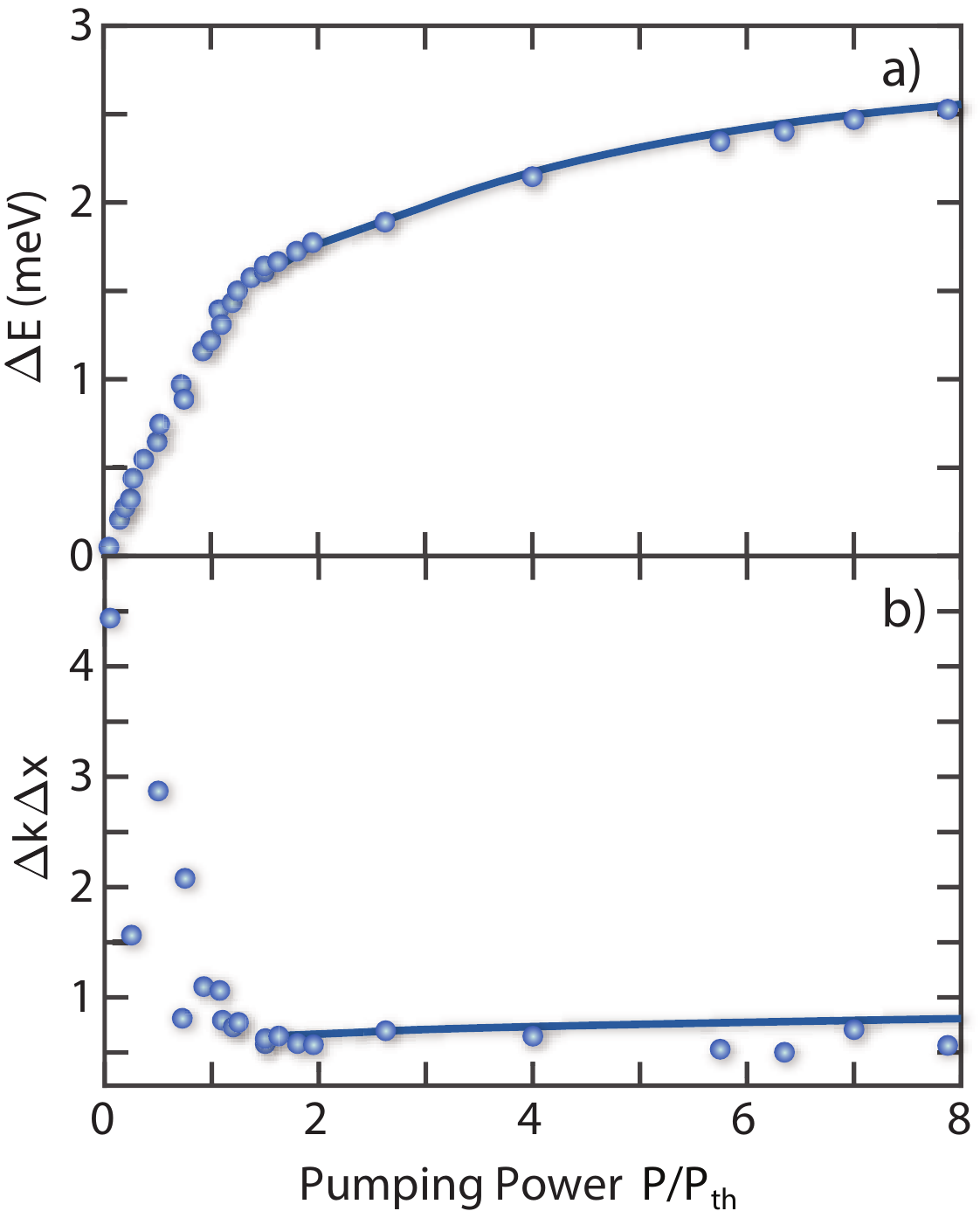}
  \caption{
    \textbf{a} The measured (dots) and calculated (lines) blue shift of the condensate energy as a function of the pump power (FWHM of the laser spot is $\SI{20}{\micro\meter}$). \textbf{b} The standard deviations $\Delta x$ and $\Delta k$ are obtained by fitting  the PL intensity of the polariton condensate in real-space and reciprocal space with a Gaussian function (see Methods). Above threshold, the  localised polariton condensate achieves the Heisenberg limit ($\Delta x \Delta k =0.5 $). The value of $\Delta x\Delta k$ as a function of power calculated  numerically is shown by the blue line (above threshold).}
\label{fig:5}
\end{figure}

In Fig.~\ref{fig:5}a, the energy blueshift of the polariton condensate is measured for increasing pumping powers and compared to the results of simulations (above threshold). The formation of the polaron-like self-trapped state is accompanied by the change  in the slope of the dependence of blueshift on the pump power visible at $P=P_{th}$ in Fig.~\ref{fig:5}a, with a slower increase above threshold. For excitation powers $P>P_{th}$, the localisation in real space corresponds to a delocalised emission in the reciprocal space (see Fig.~\ref{fig:1}), with the product of the standard deviations of  the position and momentum distributions approaching the limit imposed by the Heisenberg uncertainty principle, $\Delta x \Delta k = 0.5$ (Fig.~\ref{fig:5}b). This shows that the polaron self-trapping mechanism studied here results in the formation of the strongest possible localised polariton condensate with the widest possible distribution in $k$-space. The range of $k$-states available at the energy of the condensate is limited by the two bright spot visible in Fig.~\ref{fig:1}e.

In conclusion, the self-trapped stationary state of bosonic condensates of exciton polaritons can be referred to as a bosonic polaron phase. Indeed, the attractive interaction of polaritons is governed by their coupling with the crystal lattice that is a signature of a polaron. In contrast to conventional Pekar polarons \cite{Landau1948} the dressing of a polariton condensate is produced by the local heating of the lattice rather than by its mechanical deformation. A number of polaritons contributing to the bosonic polaron is estimated as $2\times 10^{3}$ (see Fig.~\ref{fig:1}g). We estimate the local lattice temperature variation  in the center of the polaron state as being of the order of $\SI{10}{K}$ (see Methods). The strong dependence of the shape of a polaron state on the geometry of the pumping light beam offers an opportunity for efficient optical manipulation of polaron states that is crucial for their applications in polariton simulators.

\begin{acknowledgements} Funding from the POLAFLOW ERC Starting Grant, the ERC project ElecOpteR grant number 780757, Beyond-Nano National Operative Programm PON project, Polatom ESF Research Networking Program is acknowledged. 
\end{acknowledgements}


\section*{Methods}
\subsection*{Experiment}
The semiconductor device used in these experiments is a high quality-factor ($Q>10^{5}$) $3/2\lambda$ planar cavity with 12 GaAs quantum wells (width=7nm) placed at the three electric field antinode positions. Front and back mirrors consist  of 34 and 40 $AlAs/Al_{0.2}Ga_{0.8}As$ pair layers, respectively. The light-matter interaction, measured by the Rabi splitting (energy separation between the upper and lower polariton branches at the resonant exciton-cavity condition) is of $\SI{15}{\milli\electronvolt}$. We choose a sample surface area in which the detuning between the photon and exciton modes is about $\delta=\SI{-3}{\milli\electronvolt}$. The polariton condensation is achieved under non resonant pumping using a single mode Ti:sapphire laser (excitation wavelength at the first minimum of the stop band, $\approx735$ nm, polariton emission at 774 nm).

The sample is kept at a temperature of $\SI{10}{\kelvin}$ in a low-vibration cryostat, and the excitation power is chopped at a frequency of $\SI{4}{\kilo\hertz}$ with a duty cycle of 5\%, giving a measurement time window of $\approx\SI{10}{\micro\second}$, to avoid global heating of the sample while allowing for the formation of a steady-state, self-trapped condensate.

The emitted signal is collected by an objective with a numerical aperture NA=0.6 and a magnification M=25 on a CCD detector ($1024\times1344$ pixel, pixel size is $\SI{6.25}{\micro\meter}$) coupled to a monochromator giving an overall energy resolution of $\SI{60}{\micro\electronvolt}$. The momentum-space information is retrieved by imaging (M=0.1) of the reciprocal plane of the objective on the CCD detector.

Polariton density is estimated by direct calibration of the PL intensity $I$ (in Watt) using a polariton lifetime $\tau=\SI{100}{\pico\second}$ and the polariton energy $E=2\times10^{-19}\SI{}{\joule}$ following the approximate relation:
\begin{equation}
\rho=\frac{I \tau}{E}.
\end{equation}
\subsection*{Theory}
The localization of the polariton cloud in the microcavity plane is simulated with the set of coupled complex Ginzburg-Landau and rate equations which describe  an incoherently pumped exciton-polariton condensate:
\begin{subequations}\label{system}
\begin{eqnarray}
i\hbar \frac{\partial \Psi}{\partial t} &=&   \left[ -\frac{\hbar^2}{2m}\Delta_{x,y} + g_c |\Psi|^2+ g_R n -\alpha |\Psi|^2 n + \right. \\
\notag &+& \left. \frac{i \hbar}{2} \left(R n - \gamma_c \right)\right] \Psi   \label{system_a} ,\\
\frac{\partial n}{\partial t} &=& P - \left(\gamma_R + R |\Psi|^2\right)n.
\end{eqnarray}
\end{subequations}
Here $m$ is the polariton effective mass, $\Delta_{x,y}$ is the 2D Laplace operator, $g_c$ and $g_R$ are polariton-polariton and polariton-reservoir interaction parameters,  $\gamma_c$ and $\gamma_R$ are the condensate and reservoir relaxation rates respectively, $Rn$ determines the rate of the condensate replenishment by the exciton relaxation from the reservoir. The term $\alpha Rn$ in \eqref{system_a} accounts for the energy renormalization due to the heating of the structure, and it is responsible for the condensate self-trapping \cite{Chestnov2018}. We estimate the local temperature variation that corresponds to the chosen value of $\alpha=10.5\times 10^{-5}$~meV$\mu$m$^4$ as about $\SI{10}{K}$ at high pumping power for the background lattice temperature of $\SI{10}{K}$. We note that the constant $\alpha$ phenomenologically accounts  for the balance between the local heat coming from the relaxing excitons and the thermal diffusion in the microcavity structure.

The steady-state solution of Eqs.~\eqref{system}, $\Psi=\psi e^{-i\Delta E t/\hbar}$ and $\partial n/\partial t=0$,  is found using the iterative Newton-Raphson algorithm which simultaneously determines the blue shift $\Delta E$ and the order parameter $\psi(x,y)$. The Fourier transform $S(k_x,k_y)=\int \psi(x,y) e^{-ik_xx-ik_yy} dxdy$ is used to obtain the reciprocal space distribution of the condensate emission.

The following parameters have been used to fit the experimental results:
$m=0.5m_e$, where $m_e$ is free electron mass, $\gamma_c=0.01$~ps$^{-1}$,
 $\gamma_R=0.0035$~ps$^{-1}$, $\hbar R=0.072$~$\mu$eV$\mu$m$^2$,
  $g_c= 7.9$~$\mu$eV$\mu$m$^2$,
   $g_R =2g_c$.
\clearpage

\end{document}


\flushbottom

\title{Self-trapping of exciton-polariton condensates\\
Supplementary Information}

\author{Dario~Ballarini}
\affiliation{CNR NANOTEC---Institute of Nanotechnology, Via Monteroni, 73100 Lecce, Italy}

\author{Igor~Chestnov}
\affiliation{Institute of Natural Sciences, Westlake Institute for Advanced Study, Westlake University, Hangzhou, China}
\affiliation{Vladimir State University, 600000 Vladimir, Russia}

\author{Davide~Caputo}
\affiliation{CNR NANOTEC---Institute of Nanotechnology, Via Monteroni, 73100 Lecce, Italy}
\affiliation{University of Salento, Via Arnesano, 73100 Lecce, Italy}

\author{Milena~De~Giorgi}
\affiliation{CNR NANOTEC---Institute of Nanotechnology, Via Monteroni, 73100 Lecce, Italy}

\author{Lorenzo~Dominici}
\affiliation{CNR NANOTEC---Institute of Nanotechnology, Via Monteroni, 73100 Lecce, Italy}

\author{Kenneth~West}
\affiliation{PRISM, Princeton Institute for the Science and Technology of Materials, Princeton Unviversity, Princeton, NJ 08540}

\author{Loren~N.~Pfeiffer}
\affiliation{PRISM, Princeton Institute for the Science and Technology of Materials, Princeton Unviversity, Princeton, NJ 08540}

\author{Giuseppe~Gigli}
\affiliation{CNR NANOTEC---Institute of Nanotechnology, Via Monteroni, 73100 Lecce, Italy}
\affiliation{University of Salento, Via Arnesano, 73100 Lecce, Italy}

\author{Alexey~Kavokin}
\affiliation{Institute of Natural Sciences, Westlake Institute for Advanced Study, Westlake University, Hangzhou, China}

\author{Daniele~Sanvitto}
\affiliation{CNR NANOTEC---Institute of Nanotechnology, Via Monteroni, 73100 Lecce, Italy}
\affiliation{INFN, Sez. Lecce, 73100 Lecce, Italy}


\maketitle

\section*{Numerical simulations}

\begin{figure}[h]
  \includegraphics[width=0.7\linewidth]{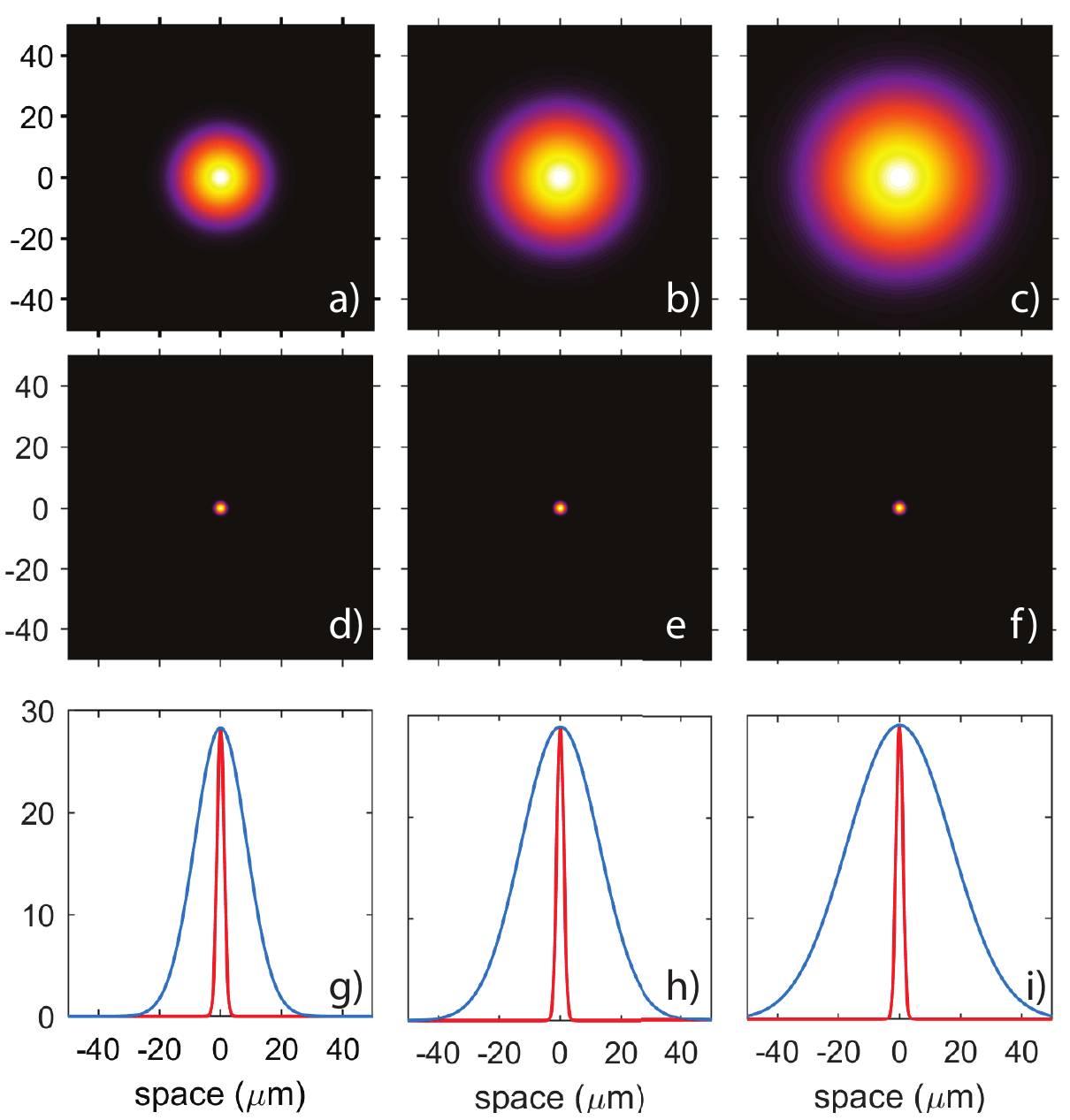}
  \caption{
    \textbf{a-f,} Calculated pump (upper row) and condensate density (middle row) distributions for different FWHM of the  symmetric Gaussian  pump: $\SI{20}{\micro\meter}$, $\SI{30}{\micro\meter}$ and $\SI{40}{\micro\meter}$ (from the left to the right). \textbf{g,h,i} Cross-sections of the intensity profiles of the pump (blue line) and the condensate (red line).}
\label{fig:4}
\end{figure}

In Fig.~\ref{fig:4}, the results of numerical simulations are shown: the polariton condensate is localised in a tight region that is much smaller than the pumped area, in perfect agreement with the experimental results as can be seen by comparing Fig.~\ref{fig:4} with Fig.~2 in the main text.
In the bottom row of Fig.~\ref{fig:4}, the cross sections of the pump and condensate intensities are plotted. Comparing them one can conclude on the localisation of the polariton condensate for different sizes of the pump spot.

\section*{Top-hat excitation}

\begin{figure}[h]
\includegraphics[width=12cm]{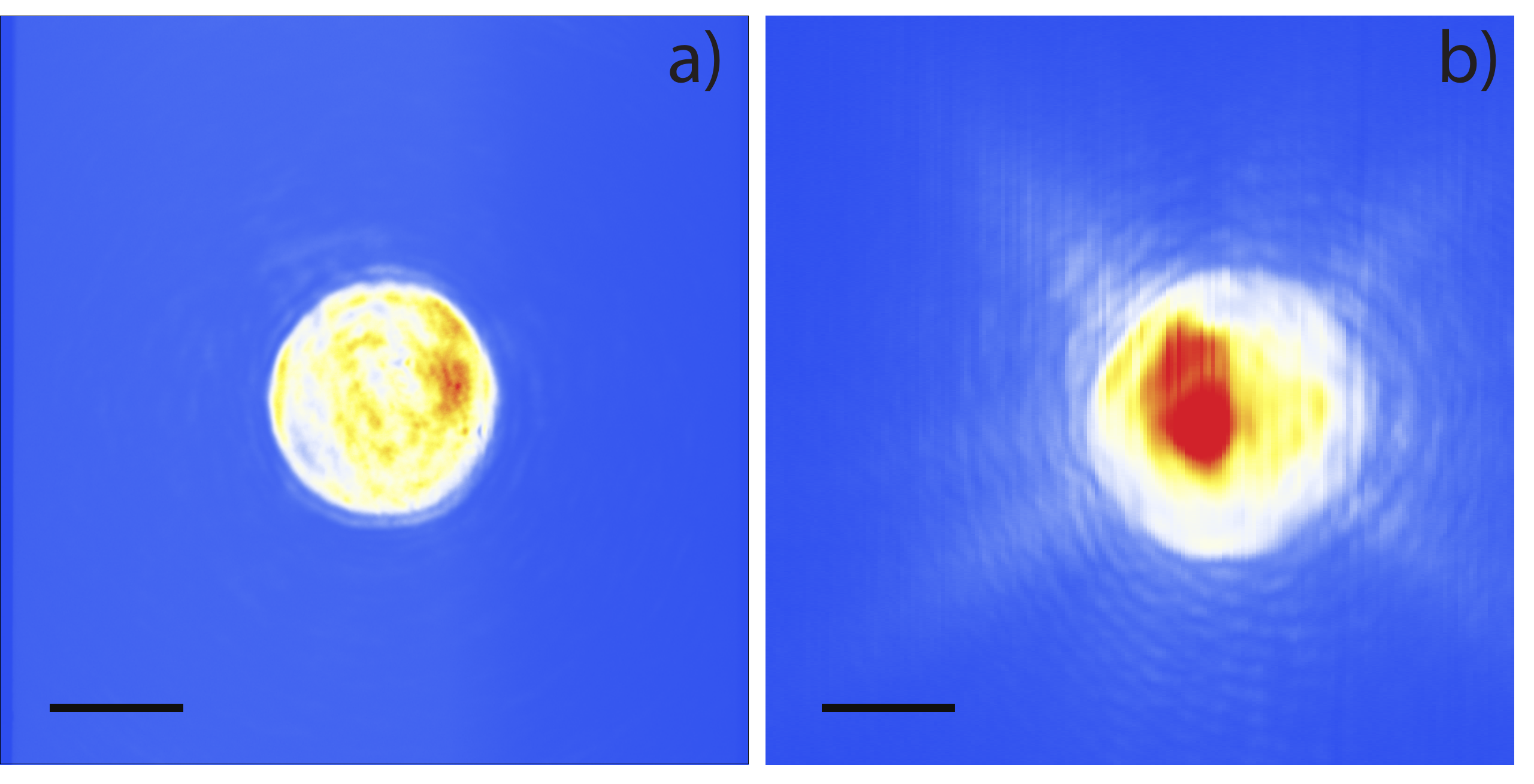} \caption{\textbf{a,} Intensity of the laser excitation spot  in 2D space. \textbf{b,} Real-space 2D distribution of the condensate. Scale bar is 40 microns.}
\label{fig:S1} 
\end{figure}

To confirm the importance of the shape of the repulsive potential created by the reservoir for the formation of a self-trapped polariton condensate, a loosely flat laser spot with a diameter of about 75 micron is used (Fig.~\ref{fig:S1}a). The condensate forms within the spot region, where the blueshift is slightly lower due to the not-perfect shape of the excitation laser spot (Fig.~\ref{fig:S1}b). The lateral size of the condensate is one order of magnitude larger than the polaron state formed on top of the Gaussian potential (see Figs.~1d,f of the main text). We conclude that in the case of excitation by a laser beam having a top-hat intensity profile the spatial size of the condensate is limited by the shallow inhomogeneities in the potential and it is not the result of the polaron effect.